# Development of the ALMA-North America Sideband-Separating SIS Mixers

Anthony R. Kerr, *Life Fellow, IEEE*, Shing-Kuo Pan, *Member, IEEE*, Stéphane M. X. Claude,
Philip Dindo, Arthur W. Lichtenberger, and Eugene F. Lauria, *Member, IEEE*.

*Abstract* — As the Atacama Large Millimeter/submillimeter Array (ALMA) nears completion, 73 dual-polarization receivers have been delivered for each of Bands 3 (84-116 GHz) and 6 (211-275 GHz). The receivers use sideband-separating superconducting Nb/Al-AlOx/Nb tunnel-junction (SIS) mixers, developed for ALMA to suppress atmospheric noise in the image band. The mixers were designed taking into account dynamic range, input return loss, and signal-to-image conversion (which can be significant in SIS mixers). Typical SSB receiver noise temperatures in Bands 3 and 6 are 30 K and 60 K, resp., and the image rejection is typically 15 dB.

*Index Terms* — Millimeter-wave, heterodyne, superconductor, sideband-separating mixer, image rejecting mixer, SIS junction, SIS receiver, radio astronomy, interferometer, ALMA.

## I. INTRODUCTION

The superconducting tunnel diode, specifically the Superconductor-Insulator-Superconductor junction, is now the preferred mixing element for low-noise heterodyne receivers at millimeter and submillimeter wavelengths. Since the introduction of the SIS mixer in 1979 [1][2][3], it has evolved from a delicate, short-lived, Pb-alloy device, difficult to fabricate reproducibly and marginally suitable for field operation, to today's relatively robust and reproducible device, usually with Nb/Al-AlOx/Nb tunnel junctions. Early SIS mixers had SIS junctions suspended across a waveguide and one or more mechanical tuners to tune out the relatively large junction capacitance [4], a configuration with limited instantaneous bandwidth and far from suitable for operation in large numbers and at remote sites, as required for the Atacama Large Millimeter/submillimeter Array (ALMA) [5]. The adaptation of photolithographic techniques from the semiconductor industry for use with Nb-based circuits has made possible the integration of tuning circuits with the SIS junctions to yield relatively wideband operation, typically 1.3:1 RF bandwidth and ~8 GHz IF bandwidth, without mechanical tuners [6][7][8].

Before ALMA, SIS receivers operated in the simple double-sideband mode, responding to signals in both upper and lower sidebands. For spectral-line radio astronomy, the main focus of ALMA, atmospheric noise entering a DSB receiver in the unwanted image sideband adds substantially to the overall system noise, thereby degrading the sensitivity of the instrument. For this reason, the North American ALMA receivers use sideband-separating SIS mixers. Sideband separation can be achieved in several ways in heterodyne receivers. An input diplexer can be used to separate the upper and lower sidebands into separate double-sideband mixers. At millimeter and submillimeter wavelengths mechanically tuned quasioptical diplexers such as the Mach-Zehnder and Martin-Puplett interferometers have been used [9][10][11]. However these have insufficient bandwidth [12] to operate with ALMA's octave or wider RF bands. For the ALMA Band-3 (84-116 GHz) and Band-6 (211-275 GHz) receivers the phasing type of sideband-separating mixer is used [13] as depicted in Fig. 1. An early design integrating the elements of a complete 200-300 GHz sideband-separating SIS mixer (RF quadrature hybrid, LO couplers, and two SIS mixers with RF tuning circuits) on a single chip was largely successful [13][14]. However, the large size of the chips would have required many wafers to be produced, with significantly higher cost than a waveguide-based circuit incorporating a pair of much smaller DSB mixer chips, each containing a single SIS mixer with its RF tuning circuit and a waveguide coupling probe [15].

## II. DESIGN OF THE ELEMENTAL SIS MIXERS

### A. SIS Mixer Theory

The quantum theory of SIS mixers, published in 1983 and 1985 by Tucker and Feldman [16][17], has been the basis for all subsequent SIS mixer design. The noise and conversion

Manuscript received XXXXXX; revised XXXXXXX; accepted XXXXXX. Date of publication XXXXXX; date of current version XXXXXX. This work was supported in part by the National Radio Astronomy Observatory. The National Radio Astronomy Observatory is a facility of the National Science Foundation operated under cooperative agreement by Associated Universities, Inc.

A. R. Kerr and S.-K. Pan are with the National Radio Astronomy Observatory, Charlottesville, VA 22903 USA (email: akerr@nrao.edu, span2@nrao.edu).

S. M. X. Claude and P. Dindo are with the National Research Council – Herzberg, Victoria, BC V9E 2E7 Canada (email: Stephane.Claude @nrc-cnrc.gc.ca, Philip.Dindo@nrc-cnrc.gc.ca).

A. W. Lichtenberger is with the University of Virginia, Charlottesville, VA 22904 USA (email: awl11@virginia.edu).

E. F. Lauria is with the University of Arizona, Tucson, AZ 85721 USA (email: glauria@email.arizona.edu).

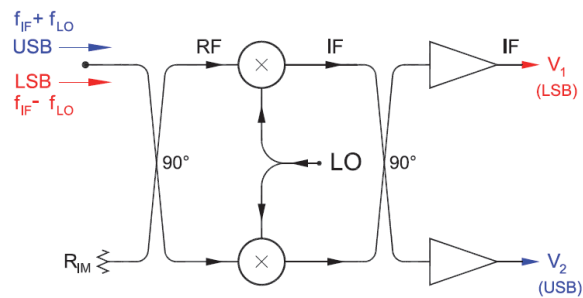

Fig. 1. The phasing type of sideband-separating mixer as used in the Band-3 and -6 ALMA SIS receivers.

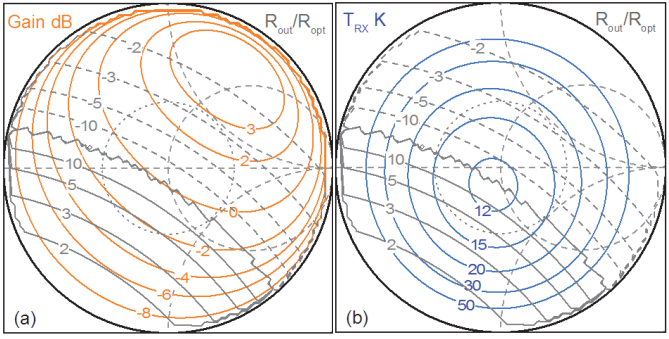

Fig. 2. For a typical Band 3 SIS receiver, contours of (a) mixer gain dB (solid orange lines) and (b) receiver noise temperature (solid blue lines), plotted on a Smith chart of the RF source impedance normalized to the optimum source impedance $R_{opt}$. The gray contours are the IF output resistance of the mixer normalized to $R_{opt}$: solid gray lines indicate positive output resistance and dashed gray lines indicate negative output resistance. The squiggly gray line indicates the transition from positive to negative output resistance. The dashed gray circles at $|\rho| = 0.4$ indicate the boundary of the desirable operating region. A 4 K IF isolator is used and an amplifier noise temperature of 4 K is assumed.

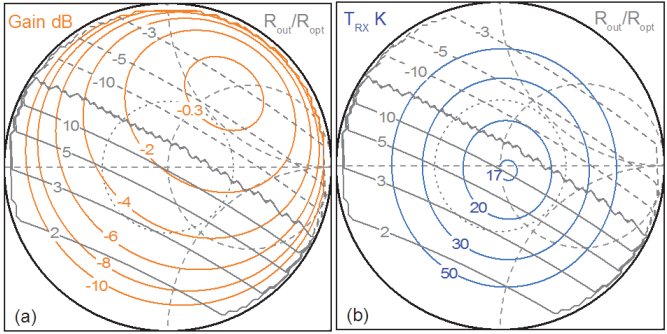

Fig. 3. For a typical Band 6 SIS receiver, contours of (a) mixer gain dB (solid orange lines) and (b) receiver noise temperature (solid blue lines), plotted on a Smith chart of the RF source impedance normalized to the optimum source impedance $R_{opt}$. The gray contours are the IF output resistance of the mixer normalized to $R_{opt}$: solid gray lines indicate positive output resistance and dashed gray lines indicate negative output resistance. The squiggly gray line indicates the transition from positive to negative output resistance. The dashed gray circles at $|\rho| = 0.4$ indicate the boundary of the desirable operating region. A 4 K IF isolator is used and an amplifier noise temperature of 4 K is assumed.

characteristics of an SIS mixer depend on the embedding impedance seen by the junction at the sideband frequencies $nf_{LO} \pm f_{IF}$ and at the LO harmonics $nf_{LO}$. Under typical operating conditions, for an SIS junction with normal resistance $R_N$, an optimum RF source impedance, based on the Tucker theory, is given by the Ke and Feldman formula [18]:

$$R_{opt} = 4R_N \left( 2 + \frac{eV_g}{hf_{LO}} \right)^{-1},$$

where $eV_g$ is the superconducting energy gap, $h$ is Planck's constant and $f_{LO}$ is the local oscillator frequency. Figs. 2-5 show how the characteristics of typical Band-3 and -6 ALMA mixers depend on the RF source impedance. Fig. 2(a) shows simulated contours of gain in dB (heavy orange lines) and IF output impedance normalized to $R_{opt}$ (gray lines, dashed where the output resistance is negative) for a Band-3 mixer biased near the middle of the first photon step and with LO power adjusted for a pumping parameter $\alpha = eV_{LO}/hf = 1.2$. Fig 2(b) shows contours of simulated receiver noise temperature $T_{RX}$. Figs. 3(a) and (b) show the same for a typical Band-6 mixer. The contours in Figs. 2 and 3 are plotted on Smith charts of the RF source impedance (which includes the junction capacitance) normalized to $R_{opt}$. In the simulations, the LO voltage waveform is assumed to be sinusoidal; this is likely to be a good approximation for SIS junctions which, by virtue of their parallel-plate construction, have relatively large shunt capacitance. The junctions are assumed to be terminated in their own capacitance at the second harmonic sideband frequencies $2f_{LO} + f_{IF}$, and short circuited by their own capacitance at higher frequencies – this is the *quasi-five frequency* case [19] which has been found to give reasonably accurate results.

It is evident in Figs. 2 and 3 that the output impedance of the SIS mixers is negative when the source impedance lies in much of the upper half of the Smith chart, and that the maximum conversion gain occurs in the region of negative output resistance. Negative resistance and gain > 1 are a result of the non-classical nature of the SIS mixer and are not parametric effects – unlike a semiconductor diode, the capacitance of an SIS junction is not voltage dependent. The implications of negative output resistance in designing SIS mixer circuits are discussed in [20]. To optimize the overall receiver noise, it can be appropriate to operate an SIS mixer in the region of high output resistance, positive or negative. To avoid a large RF input mismatch, and even negative RF input resistance, the IF load impedance should not be large relative to $R_{opt}$. We have found that an IF load equal to the optimum RF source resistance $R_{opt}$ is a good compromise. This also results in acceptable signal-to-image conversion.

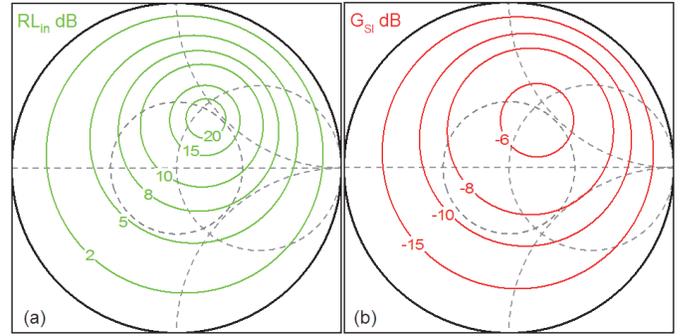

Fig. 4. For a typical Band 3 SIS receiver, contours of (a) RF input return loss dB (solid green lines) and (b) signal-to-image conversion gain dB (solid red lines), plotted on a Smith chart of the RF source impedance normalized to the optimum source impedance $R_{opt}$. The dashed gray circles at $|\rho| = 0.4$ indicate the boundary of the desirable operating region.

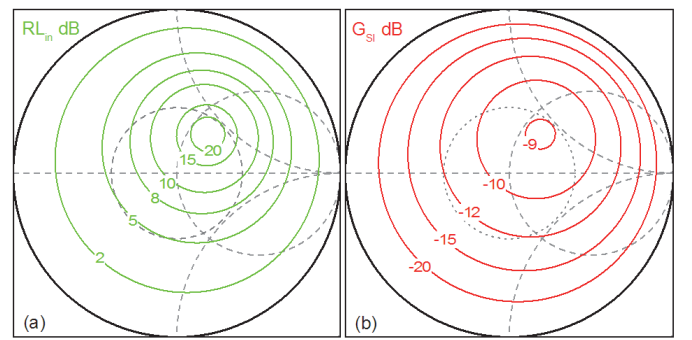

Fig. 5. For a typical Band 6 SIS receiver, contours of (a) RF input return loss dB (solid green lines) and (b) signal-to-image conversion gain dB (solid red lines), plotted on a Smith chart of the RF source impedance normalized to the optimum source impedance $R_{opt}$. The dashed gray circles $|\rho| = 0.4$ indicate the boundary of the desirable operating region.

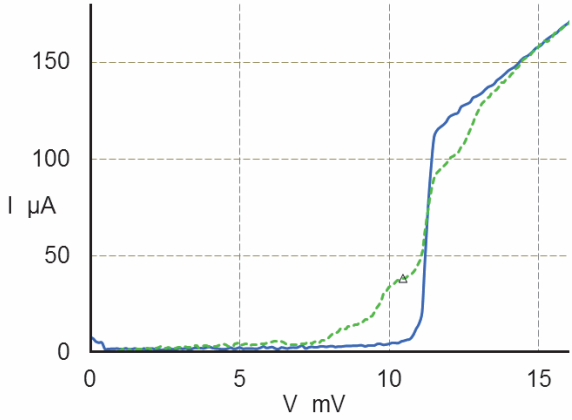

Fig. 6. Pumped and unpumped I(V) curves of the a typical four-junction SIS array used in the Band-3 mixers. The operating bias point is indicated by the marker.

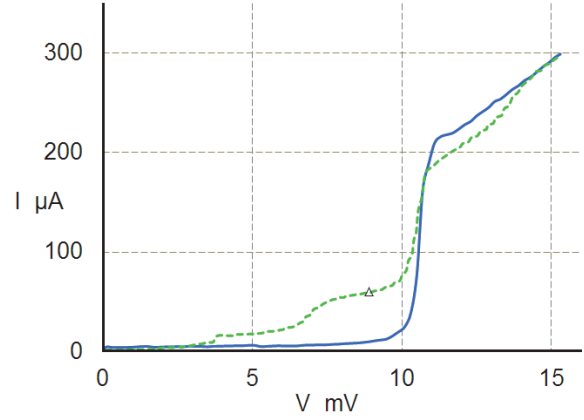

Fig. 7. Pumped and unpumped I(V) curves of the a typical four-junction SIS array used in the Band-6 mixers. The operating bias point is indicated by the marker.

The signal-to-image conversion gain, $G_{SI}$, is particularly important in sideband-separating receivers because signal power converted to the image frequency and then partially reflected by any mismatched components in the input circuit, such as an imperfect OMT or feed horn, will appear at IF as a spurious image frequency signal. Figs. 4 and 5 show contours of input return loss and signal-to-image gain for Bands 3 and 6, plotted on a Smith chart of the RF source impedance normalized to the optimum RF source resistance $R_{opt}$ as in Figs. 2 and 3.

Bands 3 and 6 use series arrays of four SIS junctions with an optimum source resistance near 50 ohms, which is a convenient impedance level for coupling to the waveguide probe and allows direct connection to a 50-ohm IF amplifier. The use of a series array of $N$ junctions in an SIS mixer has two advantages compared with a single junction. For the same overall impedance level, the individual junctions of the array are larger, which makes them easier to fabricate consistently. Also, the dynamic range (saturation power) of an SIS mixer, which depends on $N^2 f^2$ [21][22], is greater; this is particularly important for the ALMA receivers which are required to have less than 5% gain compression when connected to a 373-K calibration source. It was shown theoretically by Feldman and Rudner [23] that the noise of an SIS mixer with $N$ junctions in series is the same as that of a single-junction mixer with the same impedance level as the array – i.e., for the same critical current density, the junctions in the array have $N$ times the area of the single junction. The array requires $N^2$ times the LO power, but that is not a significant limitation for current SIS receivers. The SIS junction I(V) characteristics used in the simulations, are shown in Figs. 6 and 7.

The mixing properties of an SIS junction are largely defined by its normal resistance $R_N$, which depends on the junction area $A$ and critical current density $J_C$ [24], and by the junction capacitance $C_J$. For the Nb/Al-AlOx/Nb junctions used in this work, $R_N A = 1.8 \times 10^{-3}/J_C$, and the specific junction capacitance $C_S$ is given by the empirical expression [25]

$$C_S = 24.9 \log_{10}(J_C(\text{A/cm}^2)) - 18.1 \quad \text{fF/μm}^2.$$

The choice of junction area and critical current density to give the desired $R_N$ requires a compromise between junction quality and reproducibility on the one hand versus junction capacitance on the other. A small junction with low capacitance will have wider RF bandwidth but requires a higher value of $J_C$ (for the desired $R_N$) which results in an I(V) characteristic with greater leakage current and therefore higher shot noise. For the ALMA Band-3 mixers, the nominal junction diameter is 2.2 μm, $J_C$ = 2,500 A/cm², $R_N$ = 76 ohms, and $C_S$ = 67 fF/μm², and for Band 6, junction diameter 1.7 μm, $J_C$ = 5,200 A/cm², $R_N$ = 61 ohms, and $C_S$ = 74 fF/μm². (The $R_N$ values are for the four junctions in series.)

### B. Circuit Design

To maximize the RF coupling bandwidth between a resistive source and a series array of capacitive devices, such as SIS junctions, the series inductance of the array must be optimized as described in [20]. If the array inductance is small it may be desirable to increase it, but too large an array inductance can severely limit the bandwidth.

A wide IF bandwidth requires the inductance in the IF circuit to be minimized. This is achieved using the configuration described in [7] in which the IF current bypasses the waveguide coupling circuit.

Substrate materials compatible with the niobium SIS mixer fabrication process are silicon ($\varepsilon_r$ = 12) and fused quartz ($\varepsilon_r$ = 3.8). A fused quartz substrate was chosen for its relatively low dielectric constant which allows a wider substrate to be used without exciting higher modes in the substrate channel.

The mixer circuit is coupled to a full-height input waveguide by a rectangular suspended-stripline probe with a return loss > 21 dB over the full waveguide band [26]. Photographs of the Band–3 and -6 mixer chips are shown in Fig. 8. The suspended stripline waveguide probe (at the left) is connected to the mixer circuit through a broadband transition to capacitively loaded coplanar waveguide [13], a CPW with periodic capacitive

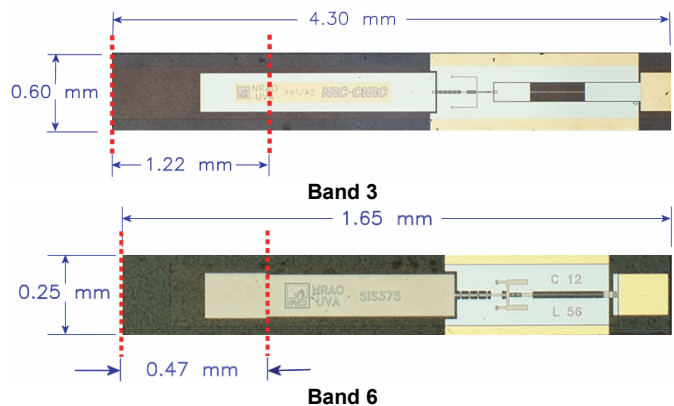

**Band 3**

**Band 6**

Fig. 8. Photographs of the Band-3 and -6 mixer chips (different scales). Dark areas are the background seen through the quartz substrate. The suspended stripline waveguide probe, to the left, connects through a broadband transition to a capacitively loaded CPW. The DC/IF bonding pad is at the right. The dotted red lines indicate the position of the broad-walls of the waveguide. Figs. 9 and 10 show more details.

loading by ground bridges to give a characteristic impedance of 50 ohms on the quartz substrate while suppressing slot modes in the ground plane.

The on-chip RF circuits of the Band-3 and -6 mixers are similar, as shown in Figs. 9 and 10. At the left is the capacitively-loaded CPW line from the suspended stripline waveguide probe, connected to a parallel pair of broadbanding resonators, the series array of four SIS junctions, a tuning capacitor $C_A$, and at the right, the end of the RF choke. For Band 3 the two resonators are quarter-wave microstrip stubs, and for Band 6 each resonator consists of a short inductive microstrip stub connected in parallel with a short capacitive stub, which approximates a parallel LC resonator connected at the input of the SIS array. The SIS mixer chips were fabricated on 50-mm diameter fused quartz wafers at the University of Virginia Microfabrication Laboratory between 2003 and 2008 using their then standard Nb/Al-oxide/Nb trilayer junction process [27].

For brevity, only the operation of the Band-6 circuit will be described in detail here. The equivalent circuit of the mixer chip is shown in Fig. 11. The different sections outlined by the dashed boxes were individually modeled using Sonnet *em* [28]. Because the conductors on the chip are all superconducting Nb, they can be considered lossless, but they have a frequency-independent surface inductance due to the London penetration depth (which is similar to the skin depth in normal conductors). The surface inductance is not of consequence in the CPW sections, but it must be taken into account in the microstrip sections in which the dielectric thickness is not much greater than the penetration depth (~85 nm in Nb). The appropriate value of surface inductance for thick and thin superconducting transmission lines is discussed in [29]. The elements in Fig. 11 were optimized using the microwave circuit simulator MMICADv2 [30]. Figs. 12 (a)-(c) show the impedance looking into the array of SIS junctions from locations A, B, and C, as indicated in Fig. 11, when the SIS junctions have an impedance equal to their optimum source impedance. The embedding impedance $Z_e$ seen by the four junctions is shown in Fig. 12 (d). Over the RF band $Z_e$ lies well within the desirable region $|\rho| \le$

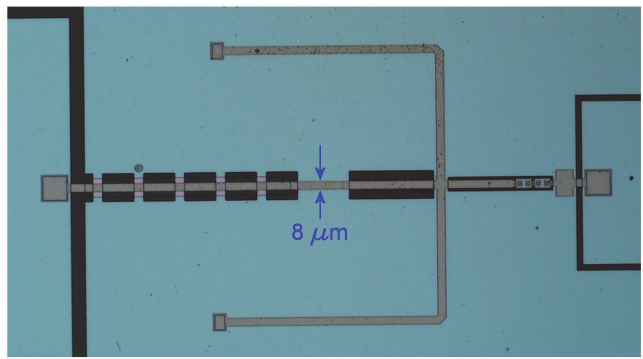

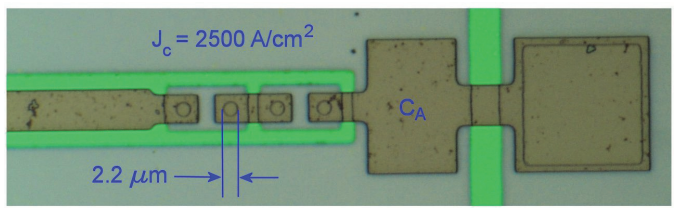

Fig. 9. RF circuit of the Band-3 SIS mixer chip. Dark or green regions are the background seen through the quartz substrate and contain no metalization. The lower Nb "base electrode" metalization is light gray; the darker (brown) conductors are the Nb "wiring" layer.

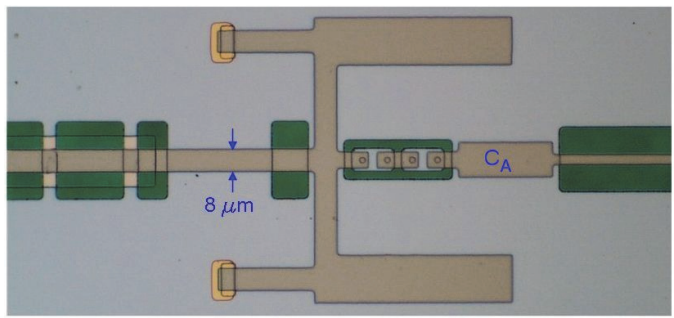

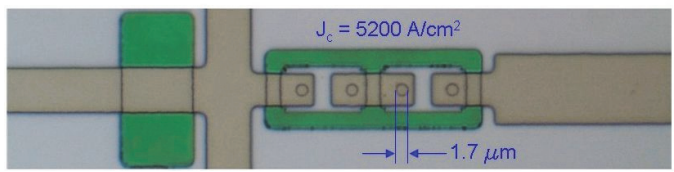

Fig.10. RF circuit of the Band-6 SIS mixer chip. Dark (green) regions are the background seen through the quartz substrate and contain no metalization. The lower Nb "base electrode" metalization is light gray; the darker (brown) conductors are the Nb "wiring" layer.

0.4. (In fact, $Z_e$ is slightly different for each of the four junctions but the differences are too small to see in this diagram.) The concept of embedding impedance needs some clarification for the case of multiple SIS junctions. It is the impedance which would be seen by a test signal source connected in place of one of the junctions when identical sources are connected in place of the other junctions. Also shown in Fig. 12(d) is the impedance seen by the junctions over the IF band, extended to 150 GHz, with the IF port of the mixer substrate terminated in $Z_{IF} = 50$ ohms . It is evident that the impedance is well behaved and less than 50 ohms over the extended IF band. This is important for maintaining stability of the mixer as explained in [20].

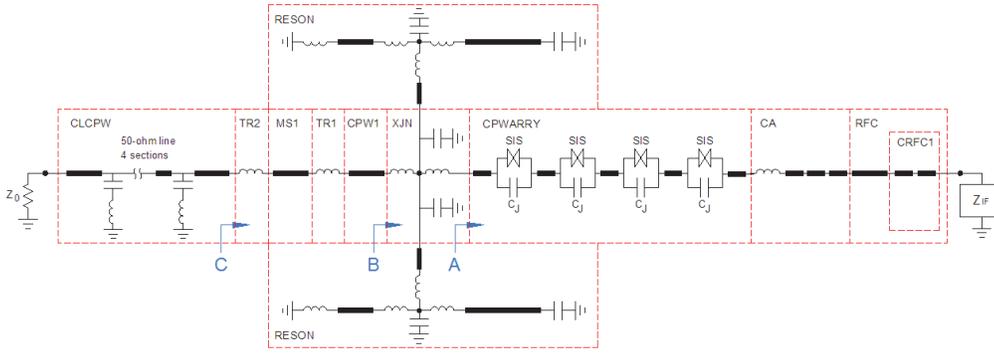

Fig. 11. Equivalent circuit of the Band-3 and Band-6 SIS mixers. At the left, the circuit connects to the suspended stripline waveguide probe. The two broadbanding resonators are at the top and bottom of the diagram (details apply to Band 6). The IF and DC connection is at the right.

## C. IF Inter-Stage Network

The IF output of the Band-3 mixer chip is connected by a pair of bond wires to a 50-ohm SMA-connector, and then by a cable to the IF isolator and amplifier. With the Band-6 mixer, no IF isolator is used and a simple phasing network between the mixer chip and the preamplifier input allows the mixer-preamp performance to be optimized for RF input match, signal-to-image conversion, and gain flatness within the IF band. Fig. 13 shows the details of the interstage network.

## D. Mixer Bias Circuit

The ALMA SIS mixers use a 6-wire bias circuit based on the one described in [31], as shown in Fig.14. The bias voltage for the mixer is developed across the 50-ohm resistor by a current source connected to $S^+$ and $S^-$, and the junction current through the 5-ohm resistor is monitored at $I_J^+$-$I_J^-$. A servo loop controls the current source to maintain a constant bias voltage at the mixer $(V_J^+$-$V_J^-)$. The presence of the 50-ohm resistor ensures that the bias circuit has a low impedance from DC to well beyond the bandwidth of the servo loop, which is desirable for maintaining the stability of the bias circuit.

For both Bands 3 and 6 the bias circuit is built into the IF preamplifier and connects to the mixers through the IF circuit, and in the case of Band 3, through the IF isolator whose center conductor is isolated from ground [32]. For Band 3 the bias circuit uses discrete components and for Band 6 it is on a custom silicon chip [33].

## E. Magnetic Circuit

Conduction in SIS junctions is by two types of carrier, superconducting Cooper pairs (of electrons), and quasi-particles (single electrons from broken Cooper pairs)[34]. It is the quasi-particle current-voltage characteristic, whose strong non-linearity is seen in Figs. 6 and 7, which is used in SIS mixers. The Cooper pair current, or Josephson current, has a complex nonlinearity which can be characterized as a voltage-controlled oscillator [35] connected in parallel with the quasi-particle circuit. The presence of the Josephson effect in an SIS mixer can seriously degrade the noise temperature [36], but fortunately the Josephson current can be suppressed by a magnetic field applied between the electrodes of an SIS junction, and most SIS mixers include a solenoid with pole pieces which concentrate a magnetic field in the vicinity of the junctions. For the Band-6 mixers a field of ~180 Gauss is produced by a current of ~26 mA in a coil of 5,500 turns of superconducting wire. In Band 3, a magnetic field has not been found necessary.

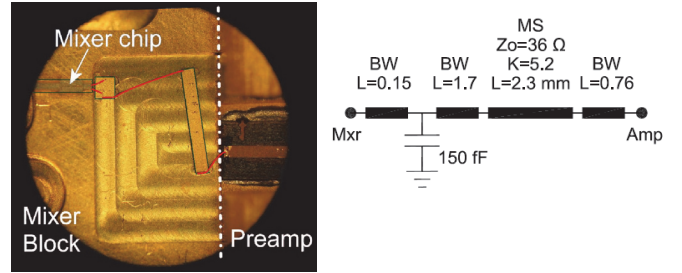

Fig. 13. Details of the IF interstage network used in the Band-6 mixer. BW = bond-wire (red), MS = microstrip. Lengths are in mm.

## III. THE SIDEBAND-SEPARATING MIXERS

For Band 6 (211-275 GHz), the complete sideband-separating mixer [37] is an E-plane split-block assembly with two gain- and phase-matched 4-12 GHz IF preamplifiers [38] attached directly to the mixer block, so IF isolators are not needed. For Band 3 (84-116 GHz), separate DSB mixer modules are attached to a split-block waveguide assembly containing the RF hybrid and LO couplers, and external 4-8 GHz cryogenic IF isolators are used [39]. Fig. 15 shows the configuration of the Band-3 and -6 mixers. An important component is the termination on the fourth port of the RF hybrid, $R_{IM}$. This is the image termination, whose thermal noise in the upper and lower sidebands is downconverted to, respectively, the lower- and upper-sideband IF outputs of the sideband-separating mixer. By including $R_{IM}$ in the cryogenic mixer assembly its contribution to the mixer noise temperature is minimized.

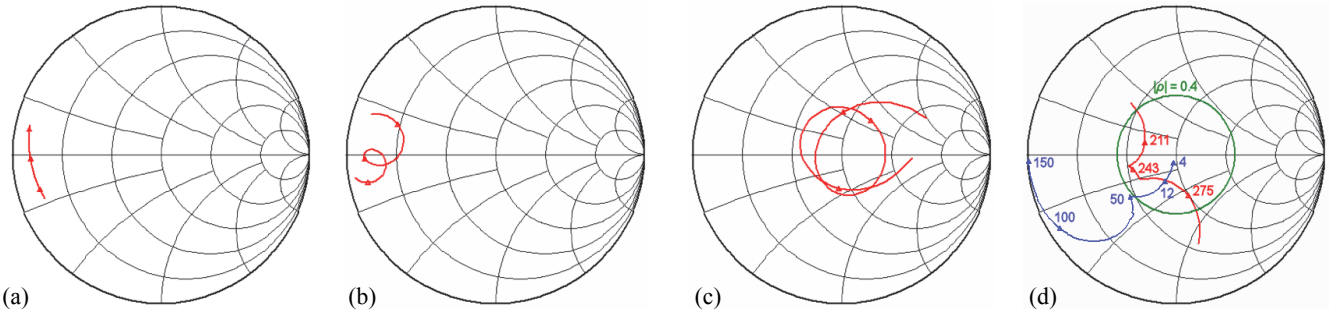

Fig. 12. Smith chart plots of the impedance at various points through the circuit of Fig. 11. (a), (b), and (c) show the reflection coefficients looking to the right at planes A, B, and C, resp., normalized to $Z_0 = 50\ \Omega$ over 200-280 GHz. (d) shows is the reflection coefficient seen by the SIS junctions over the same RF band (red), and from 4-150 GHz (blue), normalized to $R_{opt}$. The circle $|\rho| = 0.4$ (green) is included for reference.

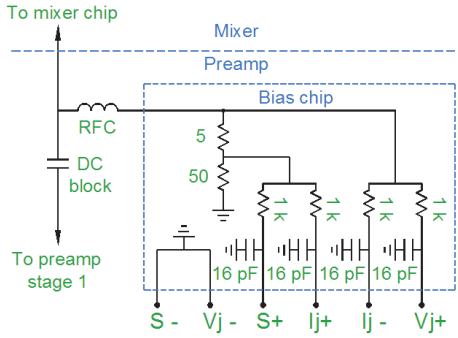

Fig. 14. Details of the SIS mixer bias circuit. The 5-ohm resistor monitors the mixer current, and the 10k resistors and 16 pF capacitors provide ESD protection.

The amplitude and phase balance of the cryogenic IF quadrature hybrid is important in ensuring good image rejection. The initial Band-6 receivers used commercial IF hybrids which were found to be inconsistent at 4 K. Subsequently a cryogenic design by Malo and Gallego [40] was implemented by MAC Technology [41] and has given excellent results.

For the Band-6 mixers, the IF preamplifiers are selected to have well matched gain and phase *vs* frequency at 4 K; this is necessary because the preamplifiers are located in the signal path before separation of the sidebands in the IF hybrid. The amplifiers were selected in pairs with gain and phase differences which would result in no less than 17 dB image rejection (see Fig. 2 of [42]). For the Band-3 configuration, matching of the preamps is not necessary because they are after the IF hybrid, but isolators are necessary to ensure that reflection of the IF signal at the amplifier inputs does not degrade the image rejection.

Fig. 16 shows the Band-6 mixer with the lid removed. It includes the branch-line waveguide RF hybrid, LO power divider and LO couplers with their fourth-port terminations, and the image termination. Fig. 17 shows the Band 6 mixer-preamp assembly (a) with the mixer and amplifier lids removed, and (b) with the IF hybrid and magnet circuit in place. Fig. 18 shows the Band 3 mixer assembly with the lids removed, and Fig. 19 shows the Band 3 mixer connected to the IF hybrid and the IF isolators and preamplifiers.

## IV. MEASUREMENT OF THE MIXERS

The noise temperature and gain of the mixer-preamps were measured using room temperature and liquid nitrogen loads placed in front of the vacuum window of a test dewar. For sideband-separating receivers it is possible to measure the image rejection accurately without knowing the power level of the RF test signal by using the procedure described in [43]. Knowing the image rejection, the SSB noise temperatures can be corrected to remove the contribution to the Y-factor from the image band (which acts to increase the Y-factor thereby reducing the apparent SSB noise temperature). The mixer-preamp gain is determined from the change in IF output power of the test receiver when RF hot and cold loads are placed in front of the receiver, and the known gain and noise temperature of the subsequent IF stages. For Band 3, the measured mixer-preamp SSB noise temperature, gain, and image rejection are shown in Fig. 20 for superior and typical mixer-preamps. The LO was tuned in 4 GHz increments, and at each LO frequency the mixer-preamp was measured across the IF band in 100 MHz steps. Similar results for Band 6 are shown in Fig. 21. The results in Figs. 20 and 21 include the contributions of the RF (quasi-optical) vacuum window, infrared filter, and cold optics in the dewar.

## V. CONCLUSION

A total of 146 Band-3 mixers have been built and tested at NRC-Canada and delivered to ALMA in 73 dual-polarized cartridges. During the 2003-2009 production period, cryogenic tests were done on 515 Band-3 DSB mixer modules and 293 sideband-separating mixer assemblies. For Band 6, a total of 158 mixer-preamps have been built and tested at the NRAO and delivered to ALMA, 146 in 73 dual-polarized cartridges and 12 spare units. During the 2008-2012 production period, cryogenic tests were done on 780 Band-6 mixer-preamps.

The SIS mixer chips for Bands 3 and 6 are all made at the University of Virginia Microfabrication Laboratory.

Additional details of the North American ALMA receivers will

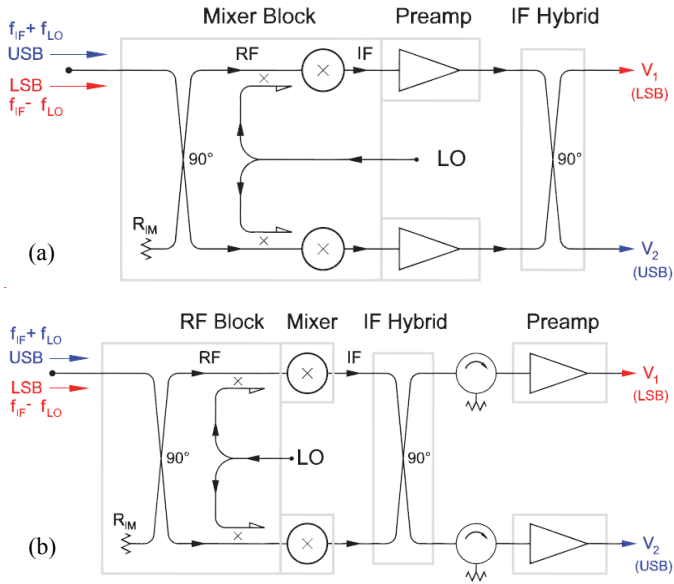

Fig. 15. Sideband-separating mixer-preamplifier configuration. (a) For Band 6, the IF preamplifiers are integrated with the mixers and the IF hybrid follows the amplifiers with no isolators. (b) For Band 3, the IF hybrid follows the mixers and the IF amplifiers are preceded by isolators.

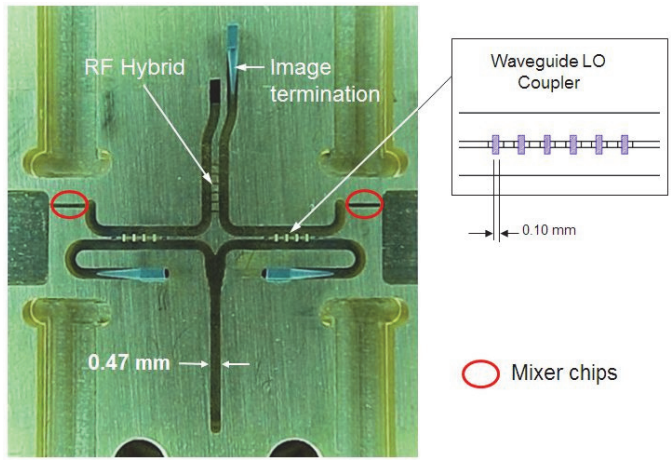

Fig. 16. One half of the Band-6 mixer, including the waveguide branch-line hybrid, LO power divider and LO couplers with their fourth-port terminations, and the image termination. The RF signal enters through a waveguide perpendicular to the page in the upper part of the circuit and the LO enters through a similar waveguide in the mating half of the block.

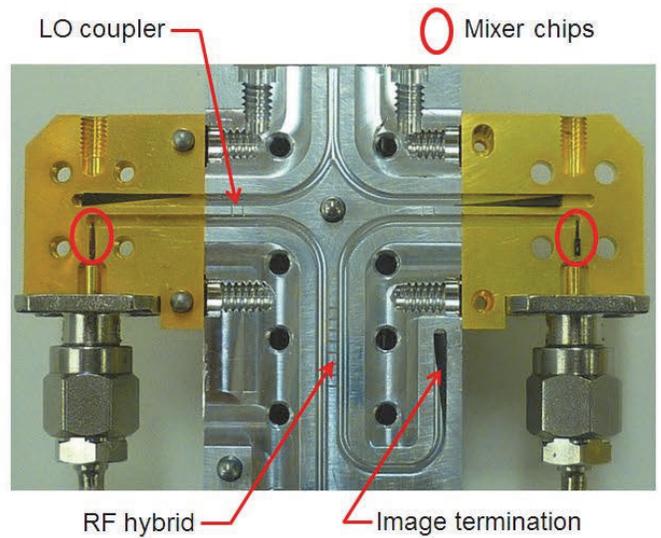

Fig. 18. One half of the Band-3 mixer assembly. The central aluminum block contains the waveguide branch-line hybrid, LO power divider, LO couplers, and the image termination. The mixer chips and LO terminations are in the left and right modules. The RF signal enters through the lower waveguide and the LO power through the upper waveguide.

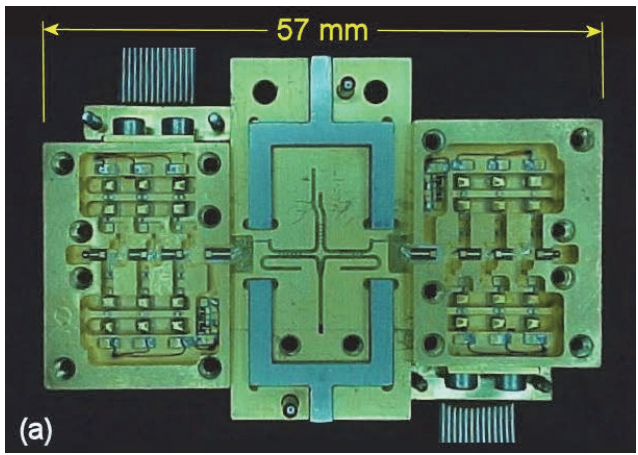

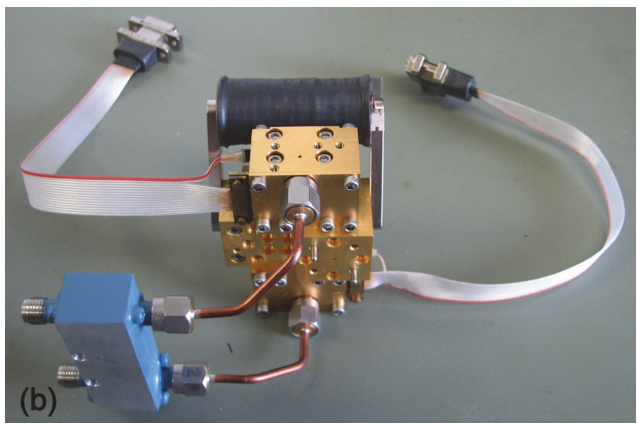

Fig. 17. Band-6 mixer assembly. (a) With preamps attached, lids removed. (b) With IF hybrid and magnet circuit.

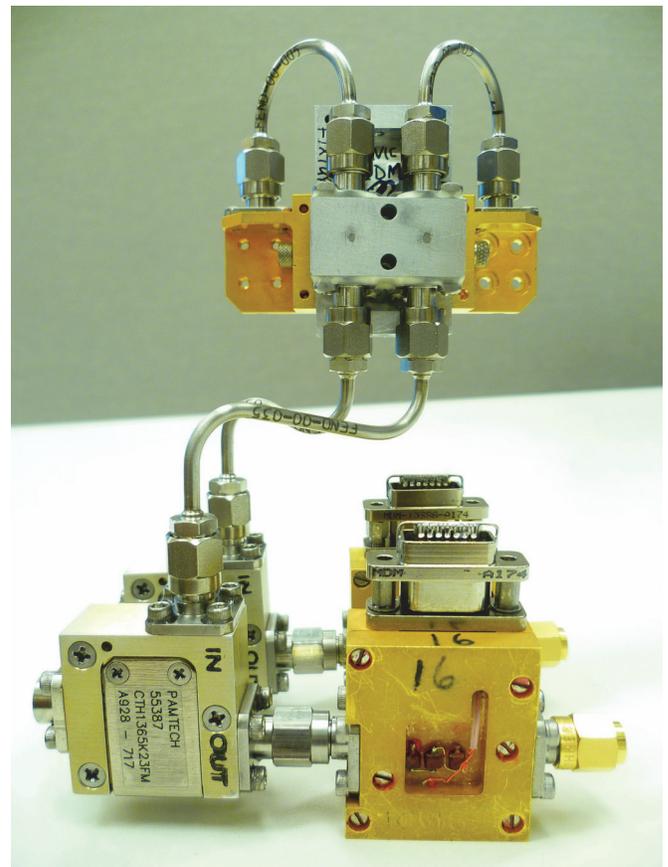

Fig. 19. Band-3 front-end. The mixer assembly is at the top, behind the IF quadrature hybrid (aluminum). Below them are the IF isolators and preamplifiers.

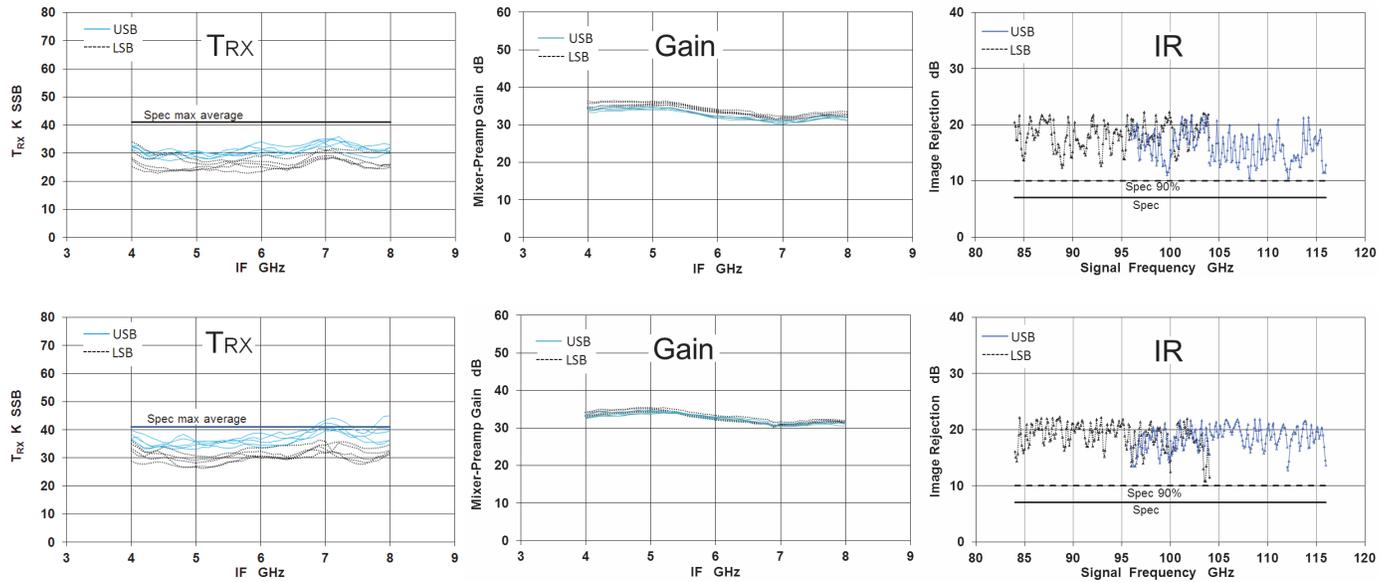

Fig. 20. Measured SSB noise temperature, gain, and image rejection, for two Band-3 mixer-preamps. The upper and lower plots are for superior and typical units. The LO was stepped from 92 to 108 GHz in 4 GHz increments at each of which the mixers were measured from over 4-8 GHz IF in 100 MHz increments. The line labeled "Spec max average" indicates the specified maximum noise temperature measured in a 4 GHz IF bandwidth and averaged over both upper and lower sidebands in both polarization channels.

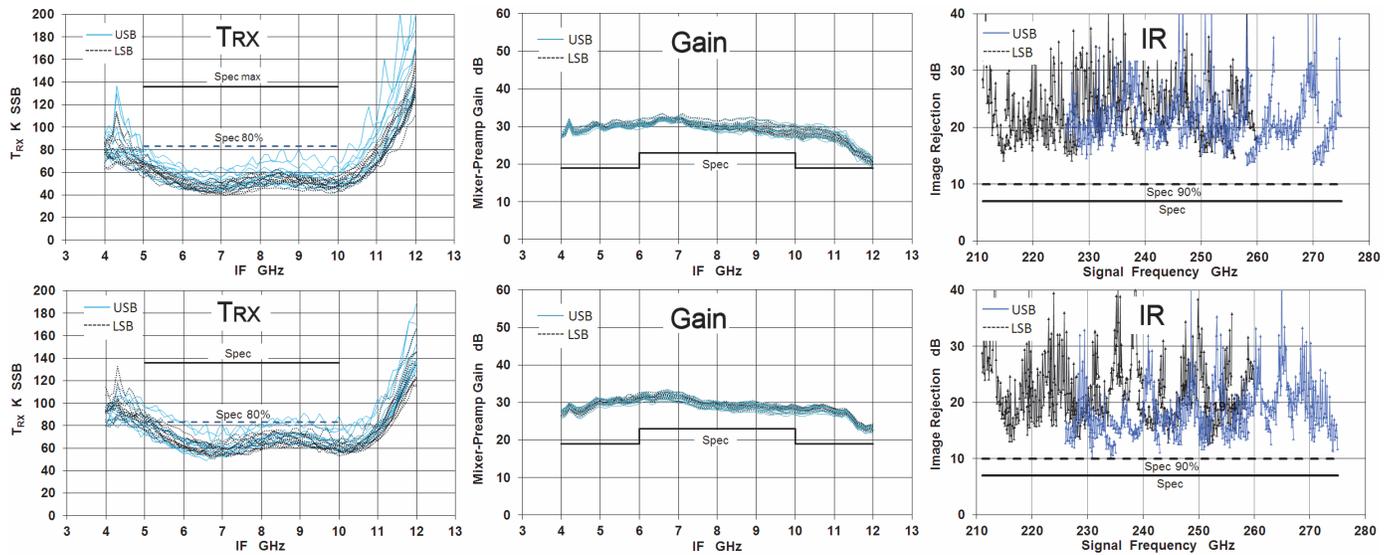

Fig. 21. Measured SSB noise temperature, gain, and image rejection, for two Band-6 mixer-preamps. The upper and lower plots are for superior and typical units. The LO was stepped from 221 to 265 GHz in 4 GHz increments at each of which the mixers were measured from over 4-12 GHz IF in 100 MHz increments. The line labeled "Spec max" is the specified maximum noise temperature, and the line "Spec 80%" is the specified maximum noise temperature over at least 80% of the band.

be found in companion papers on the Band 3 and 6 receiver cartridges[44], local oscillators[45], the photonic local oscillator phase reference system[46], and an overview of ALMA [47].

## ACKNOWLEDGMENTS

The authors gratefully acknowledge the substantial contributions of the following to the fabrication and testing of the Band-3 and -6 mixer preamps: Eric Bryerton, Dennis Derdall, John Effland, Darren Erickson, Dominic Garcia, Ralph Groves, Neil Horner, Françoise Johnson, Mike Lambeth, Tony Marshall, Matt Morgan, Greg Morris, Gerry Petencin, Marian Pospieszalski, Greg Rodrigues, Kamaljeet Saini, Dave Schmitt, and Keith Yeung and Jian-Zhong Zhang. We thank Charles Cunningham and John Webber for their enlightened management of the ALMA-North America front-end development and production.

ALMA is a partnership of ESO (representing its member states), NSF (USA) and NINS (Japan), together with NRC (Canada) and NSC and ASIAA (Taiwan), in cooperation with the Republic of Chile. The Joint ALMA Observatory is operated by ESO, AUI/NRAO and NAOJ.

## REFERENCES

[1] P. L. Richards, T. M. Shen, R. E. Harris, and F. L. Lloyd, "Quasiparticle heterodyne mixing in SIS tunnel junctions," *Appl. Phys. Lett.*, vol. 34, no. 5, pp. 345-347, 1 March 1979.
http://ieeexplore.ieee.org/stamp/stamp.jsp?tp=&arnumber=4848034

[2] G. J. Dolan, T. G. Phillips, and D. P. Woody, "Low-noise 115-GHz mixing in superconducting oxide-barrier tunnel junctions," *Appl. Phys. Lett.*, vol. 34, no. 5, pp. 347-349, 1 March 1979.
http://ieeexplore.ieee.org/stamp/stamp.jsp?tp=&arnumber=4848035

[3] S. Rudner and T. Claeson, "Arrays of Superconducting Tunnel Junctions as Low Noise 10-GHz Mixers," *Appl. Phys. Lett.*, vol. 34, no. 10, pp. 711-713, 15 May 1979.
http://ieeexplore.ieee.org/stamp/stamp.jsp?tp=&arnumber=4847872

[4] S.-K. Pan, M. J. Feldman, A.R. Kerr, and P. Timbie "Low-noise 115-GHz receiver using superconducting tunnel junctions," *Appl. Phys. Lett.*, vol. 43, no. 8, pp. 786-788, 15 Oct. 1983.
http://ieeexplore.ieee.org/stamp/stamp.jsp?tp=&arnumber=4852324

[5] A. Wootten and A. R. Thompson, "The Atacama Large Millimeter/Submillimeter Array," *Proc. IEEE*, vol. 97 , no. 8, pp. 1463 - 1471, Aug. 2009.
http://ieeexplore.ieee.org/stamp/stamp.jsp?tp=&arnumber=5136193

[6] R. Blundell, C.-Y. E. Tong, D. C. Papa, R. L. Leombruno, X. Zhang, S. Paine, J. Stern, H. G. LeDuc, and B. Bumble, "A wideband fixed-tuned SIS receiver for 200 GHz operation," *IEEE Trans. Microwave Theory Tech.*, vol. MTT-43, no. 4, pp. 933-937, April 1995.
http://ieeexplore.ieee.org/stamp/stamp.jsp?tp=&arnumber=375257

[7] A. R. Kerr, S.-K. Pan, A. W. Lichtenberger and H. H. Huang, "A Tunerless SIS mixer for 200–280 GHz with low output capacitance and inductance," *Proceedings of the Ninth International Symposium on Space Terahertz Technology*, pp. 195-203, 17-19 March 1998.
http://www.nrao.edu/meetings/isstt/papers/1998/1998195203.pdf

[8] S.-K. Pan, A. R. Kerr, M. W. Pospieszalski, E. F. Lauria, W. K. Crady, N. Horner, Jr., S. Srikanth, E. Bryerton, K. Saini, S. M. X. Claude, C. C. Chin, P. Dindo, G. Rodrigues, D. Derdall, J. Z. Zhang and A. W. Lichtenberger, "A Fixed-Tuned SIS Mixer with Ultra-Wide-Band IF and Quantum-Limited Sensitivity for ALMA Band 3 (84-116 GHz) Receivers," *Proceedings of the Fifteenth International Symposium on Space Terahertz Technology*, Northhampton, MA, April 27 -29, 2004.
http://www.nrao.edu/meetings/isstt/papers/2004/2004062069.pdf

[9] N. R. Erickson, "A Very Low-Noise Single-Sideband Receiver for 200-260 GHz," *IEEE Trans. Microwave Theory Tech.*, vol. MTT-33, no. 11, pp. 1179-1188, Nov. 1985.
http://ieeexplore.ieee.org/stamp/stamp.jsp?tp=&arnumber=1133191

[10] J. M. Payne, "Millimeter and submillimeter wavelength radio astronomy," *Proc. IEEE*, vol. 77, no. 7, pp. 993-1017, July 1989.
http://ieeexplore.ieee.org/stamp/stamp.jsp?tp=&arnumber=30751

[11] J. M. Payne, J. W. Lamb, J. G. Cochran and N. Bailey, "A New Generation of SIS Receivers for Millimeter-Wave Radio Astronomy," *Proc. IEEE*, vol. 82, no. 5, pp. 811-823, May 1994.
http://ieeexplore.ieee.org/stamp/stamp.jsp?tp=&arnumber=284748

[12] A. R. Kerr, "Image frequency suppression on the MMA," Millimeter Array Memorandum #70, National Radio Astronomy Observatory, Charlottesville VA, Dec. 1991.
http://legacy.nrao.edu/alma/memos/html-memos/alma070/memo070.pdf

[13] A. R. Kerr and S.-K. Pan, "Design of planar image-separating and balanced SIS mixers," *Proceedings of the Seventh International Symposium on Space Terahertz Technology*, pp. 207-219, 12-14 March 1996. http://www.nrao.edu/meetings/isstt/papers/1996/1996207219.pdf

[14] A. R. Kerr, S.-K. Pan, and H. G. LeDuc, "An integrated sideband separating SIS mixer for 200-280 GHz," *Proceedings of the Ninth International Symposium on Space Terahertz Technology*, pp. 215-221, 17-19 March 1998.
http://www.nrao.edu/meetings/isstt/papers/1998/1998215221.pdf

[15] S. M. X. Claude, C. T. Cunningham, A. R. Kerr and S.-K. Pan, "Design of a Sideband-Separating Balanced SIS Mixer Based on Waveguide Hybrids," ALMA Memo No. 316, 16 Aug 2000.
http://www.alma.nrao.edu/memos/html-memos/alma316/memo316.pdf

[16] J. R. Tucker, "The quantum response of nonlinear tunnel junction as detectors and mixers," Reviews of Infrared & Millimeter Waves, (Plenum, New York), vol. 1, p. 47-75, 1983.

[17] J. R. Tucker and M. J. Feldman, "Quantum detection at millimeter wavelengths," *Rev. Mod. Phys.*, vol. 57, no. 4, pp. 1055-1113, Oct. 1985.

[18] Q. Ke and M. J. Feldman, "Optimum source conductance for high frequency superconducting quasiparticle receivers," *IEEE Trans. Microwave Theory Tech.*, vol. 41, no. 4, pp. 600-604, April 1993.
http://ieeexplore.ieee.org/stamp/stamp.jsp?tp=&arnumber=231652

[19] A. R. Kerr, S.-K. Pan, and S. Withington, "Embedding impedance Approximations in the Analysis of SIS Mixers," *IEEE Trans. Microwave Theory Tech.*, vol. 41, no. 4, pp. 590-594, April 1993.
http://ieeexplore.ieee.org/stamp/stamp.jsp?tp=&arnumber=231650

[20] A. R. Kerr, "Some fundamental and practical limits on broadband matching to capacitive devices, and the implications for SIS mixer design," *IEEE Trans. Microwave Theory Tech.*, vol. MTT-43, no. 1, pp. 2-13, Jan. 1995.
http://ieeexplore.ieee.org/stamp/stamp.jsp?tp=&arnumber=363015

[21] M. J. Feldman, S.-K. Pan, and A.R. Kerr, "Saturation of the SIS mixer," *International Superconductivity Electronics Conference*, Tokyo, Digest of Technical Papers, pp. 290-292, Aug. 1987.
http://www.nrao.edu/library/Memos/Misc/Feldman_Saturation_SIS.pdf

[22] A. R. Kerr, "Saturation by Noise and CW Signals in SIS Mixers," *Proc. 13th International Symposium on Space Terahertz Technology*, Harvard University, pp. 11-22, 26-28 Mar 2002.
http://www.nrao.edu/meetings/isstt/papers/2002/2002011022.pdf

[23] M. J. Feldman and S. Rudner, "Mixing with SIS arrays," *Reviews of Infrared & Millimeter Waves*, (Plenum, New York), vol. 1, p. 47-75, 1983.

[24] T. van Duzer and C. W. Turner, *Principles of superconductive devices and circuits*, second ed., New York, Prentice Hall, 1999.

[25] D. M. Lea, A. W. Lichtenberger, A. R. Kerr, S.-K. Pan, and R. F. Bradley, "On-wafer resonant structures for penetration depth and specific capacitance measurements of Nb/Al-Al2O3/Nb trilayer films," Applied Superconductivity, (poster) 1994.

[26] A. R. Kerr, "Elements for E-Plane Split-Block Waveguide Circuits," ALMA Memo 381, 1 July 2001.
http://www.alma.nrao.edu/memos/html-memos/alma381/memo381.pdf.

[27] W. Clark, J. Z. Zhang and A. W. Lichtenberger, "Ti Quadlevel Resist Process for the Fabrication of Nb SIS Junctions," IEEE Trans. Appl. Superconductivity, vol. 13, pp. 115-118, 2003.
http://ieeexplore.ieee.org/stamp/stamp.jsp?tp=&arnumber=1211555

[28] Sonnet *em* allows the frequency-independent surface inductance of a conductor to be specified, which facilitates analysis of superconducting circuits. Sonnet Software, N. Syracuse, NY 13212.

[29] A. R. Kerr, "Surface impedance of superconductors and normal conductors in EM simulators," Millimeter Array Memorandum 245, National Radio Astronomy Observatory, Charlottesville VA, Jan. 1999.
http://legacy.nrao.edu/alma/memos/html-memos/alma245/memo245.1.pdf

[30] MMICADv2 is a product of Optotek, Kanata, Ontario, Canada.

[31] S.-K. Pan, A. R. Kerr, M. J. Feldman, A. Kleinsasser, J. Stasiak, R. L. Sandstrom, and W. J. Gallagher, "A 85-116 GHz SIS receiver using inductively shunted edge-junctions," *IEEE Trans. Microwave Theory Tech.*, vol. MTT-37, no. 3, pp. 580-592, March 1989.
http://ieeexplore.ieee.org/stamp/stamp.jsp?tp=&arnumber=21631

[32] Pamtech Inc., model CTH1365K23FM. http://pamtechinc.com

[33] Mini-Systems Inc., Attleboro, MA, part number MSI-210-0518-XX.

[34] T. van Duzer and C. W. Turner, "Principles of superconductive devices and circuits," New York, Elsevier North Holland, Inc., 1981.

[35] C. A. Hamilton, "Analog Simulation of a Josephson Junction," Rev. Sci. Instrum., vol. 43, no. 3, pp. 445-447, March 1972.
http://rsi.aip.org/resource/1/rsinak/v43/i3/p445_s1

[36] M. J. Wengler, N. B. Dubash, G. Pance, R. E. Miller, "Josephson effect gain and noise in SIS mixers," IEEE Trans Microwave Theory and Tech., vol. 40, no. 5, pp. 820-826, May 1992.
http://ieeexplore.ieee.org/stamp/stamp.jsp?tp=&arnumber=137385

[37] A.R. Kerr, S.-K. Pan, E. F. Lauria, A. W. Lichtenberger, J. Zhang, M. W. Pospieszalski, N. Horner, G. A. Ediss, J. E. Effland, R. L. Groves, "The ALMA Band 6 (211-275 GHz) Sideband-Separating SIS Mixer-Preamplifier," *Proc. 15th Int. Symp. on Space THz Tech.*, Northampton, MA, pp. 55-61, April 2004.
http://www.nrao.edu/meetings/isstt/papers/2004/2004055061.pdf

[38] E. F. Lauria, A.R. Kerr, M. W. Pospieszalski, S.-K. Pan, J. E. Effland, A. W. Lichtenberger, "A 200-300 GHz SIS Mixer- Preamplifier with 8 GHz IF Bandwidth," 2001 IEEE International Microwave Symposium Digest, pp. 1645-1648, May 2001.
http://ieeexplore.ieee.org/stamp/stamp.jsp?tp=&arnumber=967220

[39] P. Dindo, S. Claude, D. Derdall, D. Henke, D. Erickson, G. Rodrigues, A.

http://rmp.aps.org/pdf/RMP/v57/i4/p1055_1

Lichtenberger, and S.-K. Pan, "Design and Characterization of Sideband-Separating SIS Mixer RF Hybrids for the Band 3 Receiver (84-116 GHz)," Joint 30th International Conference on Infrared and Millimeter Waves and the 13th International Conference on Terahertz Electronics, pp. 409-410, Sep. 2005.
http://ieeexplore.ieee.org/stamp/stamp.jsp?tp=&arnumber=1572586

[40] I. Malo, J. D. Gallego, M. C. Diez, C. Cortes, C. Briso,"Cryogenic hybrid coupler for ultra low noise Radioastronomy receiver," *2009 IEEE International Microwave Symposium Digest*, pp. 1005-1008, 2009.
http://ieeexplore.ieee.org/stamp/stamp.jsp?tp=&arnumber=5337933

[41] MAC Technology model CA7256D.

[42] S. M. X. Claude, C. T. Cunningham, A. R. Kerr and S.-K. Pan, "Design of a Sideband-Separating Balanced SIS Mixer Based on Waveguide Hybrids," ALMA Memo No. 316, 16 Aug 2000.
http://www.alma.nrao.edu/memos/html-memos/alma316/memo316.pdf

[43] A. R. Kerr, S.-K. Pan and J. E. Effland, "Sideband Calibration of Millimeter-Wave Receivers," ALMA Memo #357, 27 March 2001.
http://www.alma.nrao.edu/memos/html-memos/alma357/memo357.pdf

[44] J. E. Effland, S. M. Claude, G. Ediss, P. Dindo, N. Jiang, N. Horner, P. Niranjanan, D. Schmitt, K. Yeung, "The ALMA North American Receivers," IEEE International Microwave Symposium Digest, June 2013.

[45] E. Bryerton, K. Saini, J. Muehlberg, D. Vaselaar, D. Thacker, "Low-Noise Sub-Millimeter Wave Local Oscillators for ALMA," IEEE International Microwave Symposium Digest, June 2013.

[46] W. Shillue, W. Grammer, C. Jacques, H. Meadows, J. Castro, J. Banda, R. Treacy, Y. Masui, R. Brito, P. Huggard, B. Ellison, J. F. Cliche, S. Ayotte, A. Babin, F. Costin, C. Latrasse, F. Pelletier, M. J. Picard, M. Poulin, P. Poulin, "A High-Precision Tunable Millimeter-Wave Photonic LO Reference for the ALMA Telescope," IEEE International Microwave Symposium Digest, June 2013.

[47] J. C. Webber, "The ALMA Telescope," IEEE International Microwave Symposium Digest, June 2013.